\documentclass{pasj00}
\begin{document}
\SetRunningHead{Y. Hibi et al.}{Common Correlations between 60, 100 and 140\ $\mu$m Intensities in the Galactic Plane and Magellanic Clouds}
\Received{}%{yyyy/mm/dd}
\Accepted{}%{yyyy/mm/dd}

\title{Common Correlations between 60, 100 and 140\ $\mu$m Intensities in the Galactic Plane and Magellanic Clouds}

%%% begin:list of authors
\author{Yasunori \textsc{HIBI}, Hiroshi \textsc{SHIBAI}, Mitsunobu \textsc{KAWADA} and Takafumi \textsc{OOTSUBO}}
\affil{Graduate School of Science, Nagoya University, Furo-cho, Chikusa-ku, Nagoya 464-8602}
\email{hibi@u.phys.nagoya-u.ac.jp}
\and
\author{Hiroyuki \textsc{HIRASHITA}}
\affil{Center for Computational Sciences, University of Tsukuba, Tsukuba, Ibaraki 305-8577}
%%% end:list of authors

%%% Please use the following style in case that sorting by 
%%% affilation is impossible. 
%
% \author{%
%   D-Firstname \textsc{D-Familyname}\altaffilmark{1}
%   E-Firstname \textsc{E-Familyname}\altaffilmark{1,2}
%   and
%   F-Firstname \textsc{F-Familyname}\altaffilmark{2}}
% \altaffiltext{1}{Address of Institute}
% \email{ddddd@xxx.xxx.xx.xx}
% \email{eeeee@xxx.xxx.xx.xx}
% \altaffiltext{2}{Address of Institute}

%% `\KeyWords{}' always has to be placed before `\maketitle'.
%% Do NOT move this preamble from here!

\maketitle

\begin{abstract}
We investigate the far-infrared SED of the Galaxy and the Magellanic Clouds by using the COBE (COsmic Background Explorer) 
/ DIRBE (Diffuse InfraRed Background Experiment) ZSMA (Zodi - Subtracted Mission Average) maps at wavelengths of 60\ $\mu$m, 
100\ $\mu$m and 140\ $\mu$m. We analyze three regions: the Galactic plane region with the Galactic latitude $\vert b\vert<5^{\circ}$, 
the Large Magellanic Cloud (LMC) region, and the Small Magellanic Cloud (SMC) region. Because the dust optical depth is much smaller in 
the far-infrared than in the visible, we may observe cumulative far-infrared radiation from regions with various interstellar radiation 
field (IRSF) in a line of sight. As consequence of considering such an effect, we find a common far-infrared color correlation between the 
140 -- 100\ $\mu$m and 60 -- 100\ $\mu$m intensity ratios in all the three galaxies. Although this color correlation cannot be explained by 
any existing model, it fits very well the far-infrared color of nearby star forming galaxies. 
\end{abstract}

\section{Introduction}

Interstellar dust is heated by interstellar radiation from stars and reradiates in the far-infrared regime. If we fix the size 
distribution and composition of interstellar dust and the spectrum of interstellar radiation field (ISRF), we are in principle able to 
predict far-infrared Spectral Energy Distribution (SED). D\'esert, Boulanger \& Puget (1990) discussed the interstellar dust size 
distribution and composition which reproduce various far-infrared data and interstellar extinction curve of the Galaxy. Dwek et al. (1997) 
investigated more comprehensive interstellar dust size distribution and composition by using the data taken by the COsmic Background 
Explorer (COBE) / Diffuse InfraRed Background Experiment (DIRBE). They concluded that their models could explain far-infrared SED and 
extinction curve except far-UV extinction. D\'esert, Boulanger \& Puget used the data taken by several instruments and in several sky 
fields, while Dwek et al. only used the data taken by the COBE/DIRBE and the COBE/ Far Infrared Absolute Spectrophotometer (FIRAS) in three 
limited fields. 

On the other hand, after fixing optical properties, size distribution and composition of dust grains, a series of papers by Draine and Li 
have modeled the far-infrared SED and the extinction curve (e.g., Draine \& Li 2001). They compared their model SEDs with observational 
far-infrared data of high Galactic latitudes (the Galactic cirrus component), of two narrow regions in the Galactic plane (Li \& Draine 
2001) and of a SMC region (Li \& Draine 2002). They concluded that their model explains the far-infrared SED of the cirrus compoents and 
the SMC. However, they could not explain the far-infrared SED from two Galactic plane regions even if a mixture of two different temperature 
components is considered. The interstellar extinction curve is successfully explained by their dust model. Zubko, Dwek \& Arendt (2004) 
presented a new interstellar dust model considering the consistency with the interstellar elemental abundances. Some of their models 
simultaneously explain the far-ultraviolet to infrared absorption and diffuse infrared emission of the interstellar dust.
 
Almost all these studies treated far-infrared radiation from cirrus at high Galactic latitude or narrow regions of the Galactic plane. 
Nevertheless, the strongest far-infrared radiation comes from the Galactic plane. Shibai, Okumura \& Onaka (1999) and Okumura 
et al. (1999) dealt with far-infrared data of the Galactic plane. They investigated the color-color correlation of the COBE/DIRBE 
far-infrared data in low Galactic latitude areas with \(\vert b \vert < 5^{\circ}\). They found that there is a strong correlation 
between 60\ $\mu$m, 100\ $\mu$m and 140\ $\mu$m intensities. Nagata et al. (2002) derived an empirical relation between the 60\ $\mu$m 
-- 100\ $\mu$m flux color and the temperature of Large Grains (LGs) based on the color-color correlation in Shibai et al. (1999). 
LGs have large heat capacities and are in radiative equilibrium with ISRF. They developed a method of estimating the total far-infrared 
($\lambda> 40\ \mu$m) flux only from 60\ $\mu$m and 100\ $\mu$m fluxes. They argued that their estimation is more accurate than previous 
ones such as Helou et al. (1988).

When we observe far-infrared radiation from the Galactic plane, since the optical depth in the far-infrared is much smaller than that 
in the visible, we can detect a sum of far-infrared radiation from interstellar dust illuminated by a variety of ISRF in a line of sight. 
We call this effect ``overlap effect''. To avoid the overlap effect, previous works have excluded the Galactic plane data as can be seen 
in D\'esert, Boulanger \& Puget (1990) and Dwek et al. (1997). There are only a few studies making use of the Galactic plane data (Shibai, 
Okumura \& Onaka 1999, Okumura et al. 1999 and Nagata et al. 2002), and an overall analysis of the far-infrared data of the entire Galactic 
plane is still required to comprehend the intrinsic far-infrared emission properties of interstellar dust.

In this paper, we investigate the far-infrared emission properties of interstellar dust with the COBE/DIRBE 60\ $\mu$m, 100\ $\mu$m and 140\ 
$\mu$m Zodi-Subtracted Mission Average (ZSMA) maps. Analyzed areas are the Galactic plane, the Large Magellanic Cloud (LMC) and the Small 
Magellanic Cloud (SMC). In section 2, we describe the data and the analysis method. The resultant color-color correlation and the validity of 
our analysis are examined in section 3. We discuss the far-infrared emission properties of interstellar dust in section 4, and finally we 
conclude in section 5.

\section{Data analysis}

We analyze the COBE/DIRBE ZSMA map data, which is obtained in the COBE mission by averaging all data except the data with low confidence 
(COBE/DIRBE Explanatory Supplement: the Galactic Plane Maps, 1993), and by removing the zodiacal light. The angular resolutions of the 
COBE/DIRBE all bands are 0.7 degrees. We analyze every pixel of the COBE/DIRBE map data and do not integrate in some region. 

We analyze the 60\ $\mu$m, 100\ $\mu$m and 140\ $\mu$m data since it was demonstrated by Shibai, Okumura \& Onaka (1999) that there exists a 
strong correlation among the intensities in those three bands . The fluxes in those bands are considered to originate from the thermal 
radiation from Large Grains (LGs) and Very Small Grains (VSGs). The grain radii of LGs are larger than about 20 nm and they are in radiative 
equilibrium with ISRF (e.g., Draine \& Li 2001). VSGs, which have smaller radii, are characterized by their small optical cross sections and 
heat capacities; thus, the impact of individual photon is important for heating and the fluctuation of the grain temperature is large. This 
heating process is called stochastic heating, and stochastically heated grains temporarily achieve a temperature high enough to emit 
mid-infrared photons (Aannestad \& Kenyon 1979).

We do not use data in wavelengths other than 60\ $\mu$m, 100\ $\mu$m and 140\ $\mu$m. The flux in the COBE/DIRBE 1.25\ $\mu$m, 2.2\ $\mu$m, 
3.5\ $\mu$m and 4.9\ $\mu$m bands is dominated by stellar radiation (Dwek et al. 1997). We do not adopt those wavelengths since we study the 
optical properties of interstellar dust. As for 12\ $\mu$m and 25\ $\mu$m bands of the DIRBE, Shibai, Okumura \& Onaka (1999) showed that 
the intensities in those bands correlate less strongly with 100\ $\mu$m and 140\ $\mu$m intensities, compared with the 60\ $\mu$m intensity. 
The radiation sources of 12\ $\mu$m and 25\ $\mu$m bands are VSGs and Polycyclic Aromatic Hydrocarbons (PAHs) (Dwek et al. 1997). PAHs are 
composed of carbon and hydrogen and their sizes are smaller than those of VSGs. As a result, in addition to optical properties as VSGs, 
optical properties as hydrocarbon are revealed in PAHs. Especially in the 12\ $\mu$m bands, there are several PAH band emission lines in 
addition to near to mid- infrared continuum (Dwek et al. 1997). The PAH optical properties are dealt in other works such as Draine \& Li 
(2001) and Sakon et al. (2004).

Long-wavelength data at \(\lambda=240\ \mu{\rm m}\) are also available. Dwek et al. (1997) showed that the integrated 240\ $\mu$m intensity 
in high Galactic latitude is explained by simple extension of the 100\ $\mu$m and 140\ $\mu$m intensities radiated from LGs. Thus, the 240\ 
$\mu$m band does not provide us any independent information on the LG temperature. Instead it has information on the emissivity index 
$\beta$ (see section 2.4), which we assume to be 2 in this paper. The value of emissivity index, 2, is according to the theoretical studies, 
e.g.,  Draine \& Lee (1984), and the observational studies using the COBE/FIRAS spectral resolved data, e.g., Reach et al. (1995). 
Furthermore, the 240\ $\mu$m intensity data of each pixel do not have enough S/N. We examined the emissivity index by using the data at 100, 
140, and 240 $\mu$m by assuming a single-temperature component, and found that the emissivity index of each pixel is in the range of 2 --- 4. 
Even if the emissivity index is not 2, though each band's correction factor of COBE/DIRBE will be changed, the results of this paper are 
changed little. The variations caused by the change of the emissivity index is comparable to the uncertainties of each band intensity. The 
problem of emissivity index does influence LG temperatures of each pixel. As emissivity index is larger, the LG temperature derived from the 
140 --- 100\ $\mu$m ratio is lower. For a precise determination of the emissivity index, sub-millimeter observations may be better suited 
(e.g., Aguirre et al. 2003).

\subsection{The Galactic Plane}

For the Galactic plane, we analyze the region with $0^{\circ} \leq l< 360^{\circ}$ and with $\vert b\vert < 5^{\circ}$. But because of 
the uncertainty in the zodiacal light model, we exclude the regions whose 60\ $\mu$m intensity is less than 3\ MJy\ sr$^{-1}$. The 
excluded regions are about 10 percent of the entire Galactic plane region.

Our purpose is to study the far-infrared emission properties of interstellar dust illuminated by ISRF. When we observe far-infrared 
radiation from the Galactic plane region, because the optical depth in the far-infrared is smaller than that in the visible, it is 
highly probable that we observe interstellar dust illuminated by a variety of ISRF intensity in a line of sight. In these regions, 
the hypothesis of uniform ISRF intensity in a line of sight is not valid. We identify and exclude such regions in the Galactic 
plane by taking correlation with radio continuum map in the following way.

We use the radio continuum map at 10\ GHz shown in Handa et al. (1987) and compare it with the COBE/DIRBE ZSMA map. The radio map covers 
the Galactic longitude $l$ from \timeform{356D.0} to \timeform{56D.0} and the absolute Galactic latitude $\vert b\vert < \timeform{1D.5}$. 
We employ this radio continuum map in $\vert b\vert < \timeform{1D.0}$. We compare this map with the same region of the COBE/DIRBE ZSMA map. 
The 10\ GHz radio continuum mainly comes from supernovae remnants, planetary nebulae, and H {\sc ii} region (Hirabayashi 1974), where the 
ISRF may tend to be locally high. Such high-ISRF regions should be more compact than the general interstellar medium which is known to be 
irradiated by a lower ISRF intensity (Shibai, Okumura \& Onaka 1999). Thus, a line of sight with high radio intensity is considered to 
contain a large volume of ``normal'' ISRF intensity regions and a small volume of high ISRF intensity regions, especially in the low 
Galactic latitude because of the ``overlap effect''. We distinguish such strong radio continuum regions from the other regions in the 
COBE/DIRBE ZSMA map.

\subsection{Large Magellanic Cloud}

We adopt a disk-like region with a radius of 5$^{\circ}$ centered at ($\alpha$, $\delta$) = (\timeform{5h20m}, \timeform{-68D.5}) 
(J2000) for the LMC region. In order to subtract the Galactic far-infrared component in the LMC region, we use the outer \timeform{1D.0} 
rim of the LMC region, which is contained in the defined LMC region, to determine the Galactic contribution. We fit a linear function of 
the coordinates to \(I_{\nu}(\lambda)\) at each wavelength in the rim region. Then this functional form is applied to the inner region and 
subtracted from \(I_{\nu}(\lambda)\) of the LMC region.

Considering the uncertainty in our estimate of the Galactic far-infrared intensity, our analysis is limited to the region which matches 
the following criterion. We calculate the standard deviation ($\sigma_{\lambda}$) of the residual intensity of the LMC rim region after 
subtracting the Galactic far-infrared component ($\sigma_{60\ \mu m}$=0.23\ MJy\ sr$^{-1}$, $\sigma_{100\ \mu m}$=1.0\ MJy\ sr$^{-1}$, 
$\sigma_{140\ \mu m}$=5.4\ MJy\ sr$^{-1}$). We adopt the data of the area where the intensity of the identified LMC component at all the 
wavelengths are higher than 5$\sigma_{\lambda}$. As a result, we use 13 percent of the initially determined LMC region. The adopted region 
contains the main bar and the molecular ridge region of LMC (Fukui et al. 1999).

The angular resolution of the COBE/DIRBE (\timeform{0D.7}) corresponds to 0.6 kpc at the distance of the LMC (50 kpc). In order to avoid the 
overlap effect, we discriminate 30 Doradus (30 Dor) from the other LMC regions, since there is a large variation of ISRF intensity on small 
angular scale. We identify ``30Dor region'' as a disk-like region centered at ($\alpha$, $\delta$) = (\timeform{5h40m}, \timeform{-69D.0}) 
with a radius of \timeform{1D.0}.

\subsection{Small Magellanic Cloud}

For the SMC, we adopt a disk-like region centered at ($\alpha$, $\delta$) = (\timeform{1h00m}, \timeform{-73D.0}) with a radius of 
\timeform{3D.0}. To subtract the Galactic far-infrared components, we used the same estimation and subtraction method for the Galactic 
components as in the case of the LMC in section 2.2. In order to subtract the Galactic far-infrared component in the SMC region, we use the 
outer \timeform{0D.5} rim of the SMC region.

Estimating the uncertainly in the Galactic far-infrared components ($\sigma_{60\ \mu m}$=0.19\ MJy\ sr$^{-1}$, $\sigma_{100\ \mu m}$=0.57\ 
MJy\ sr$^{-1}$, $\sigma_{140\ \mu m}$=3.65\ MJy\ sr$^{-1}$), we select the analyzed region in the same way described as in section 2.2. As 
a result, we adopt 3 percent of the initial SMC region. The selected region corresponds to the western bar (Mizuno et al. 2001), but the 
eastern wing (Mizuno et al. 2001) is excluded after our selection.

\subsection{Color Correction}

Because the published COBE/DIRBE all bands intensities are derived by assuming a source spectrum $\nu I_{\nu} =$ constant, we need to apply 
the color correction to the COBE/DIRBE far-infrared data. The SED of the LG component is expressed with the emissivity index $\beta$ as 
\(\propto\ \nu^{\beta}B_{\nu}(T)\), where $B_{\nu}(T)$ is the Planck function. The intensities at a wavelength $\lambda$ before and after 
correction are defined as \(I(\lambda)\) and \(I_{C}(\lambda)\), respectively, and are related by the color correction factor, 
\(K_{\beta =2}(T)\), as
\begin{equation}
I_{C}( \lambda ) = \frac{I( \lambda )}{K_{\beta = 2}(T_{I(140\ \mu {\rm {\rm m}})/I(100\ \mu {\rm {\rm m}}),\beta = 2})}
\end{equation}
\(K_{\beta =2}(T)\) is derived from the table in the COBE/DIRBE ``Explanatory Supplement: the Galactic Plane Map'' by assuming that $\beta\ 
=\ 2$. We also suppose that the emissions at $\lambda$=100 and 140\ $\mu$m originate from LGs. Thus the LG temperature can be determined 
from \(I_{C}(140\ \mu {\rm m})/I_{C}(100\ \mu {\rm m})\). Since we do not know $I_{C}$ before the color correction, we use 
\(I(140\ \mu {\rm m})/I(100\ \mu {\rm m})\) to estimate the temperature necessary to obtain \(K_{\beta =2}(T)\). This temperature is denoted 
as \(T_{I(140\ \mu {\rm m})/I(100\ \mu {\rm m}), \beta =2}\). In this study, the color correction factor ranges from 0.9 to 1.1 at $\lambda 
= 100\ \mu$m and 140\ $\mu$m. If the LG temperature is lower than 30 K, most of \(I(60\ \mu {\rm m})\) comes from VSGs. Practically, the LG 
temperatures of all pixels analyzed in this investigation are lower than 30 K. 

On the other hand, we mention the color correction of \(I(60\ \mu {\rm m})\). There are many investigations about the shapes of SED around 
60\ $\mu$m (e.g., Dwek et al. 1997; Li \& Draine 2001; Zubko, Dwek \& Arendt 2004), the spectral shapes can be fitted by $I_{\nu} \propto 
\nu^{-\alpha}$ with $1 < \alpha < 3$. The COBE/DIRBE correction factors for $I_{\nu} \propto \nu^{-\alpha}$ ($1 < \alpha < 3$) are 0.91 --- 
1.00. Since these are ``modeled'' SED not ``observatianal'' SED, we do not apply the color correction to \(I(60\ \mu {\rm m})\). And the 
uncertainty in the color correction will be same or much less than 60 $\mu$m band uncertainties. 

\section{Results}
\subsection{Color-Color Diagram}

We present the color-color relation between \( I(60\ \mu {\rm m}/I_{C}(100\ \mu {\rm m})\) and \(I_{C}(140\ \mu {\rm m}/I_{C}(100\ \mu {\rm 
m})\) for the Galactic plane in figure 1a. Since the number of the data points for the Galactic plane is large, we show in figure 1b the 
contour map demonstrating the density of the data points on the diagrams. The bin sizes of the two colors are both set to be 0.1 in 
logarithmic units. We draw contour lines according to the number of data points 0.5 (to express the boxes which have only one data point), 
0.5$N$, $N$, 2$N$, 3$N$ and 5$N$ on the diagram, where $N$, $N = 32$, indicates the average number of data points contained in each bin on 
this contour map.

\begin{figure}
  \begin{center}
    \FigureFile(80mm,80mm){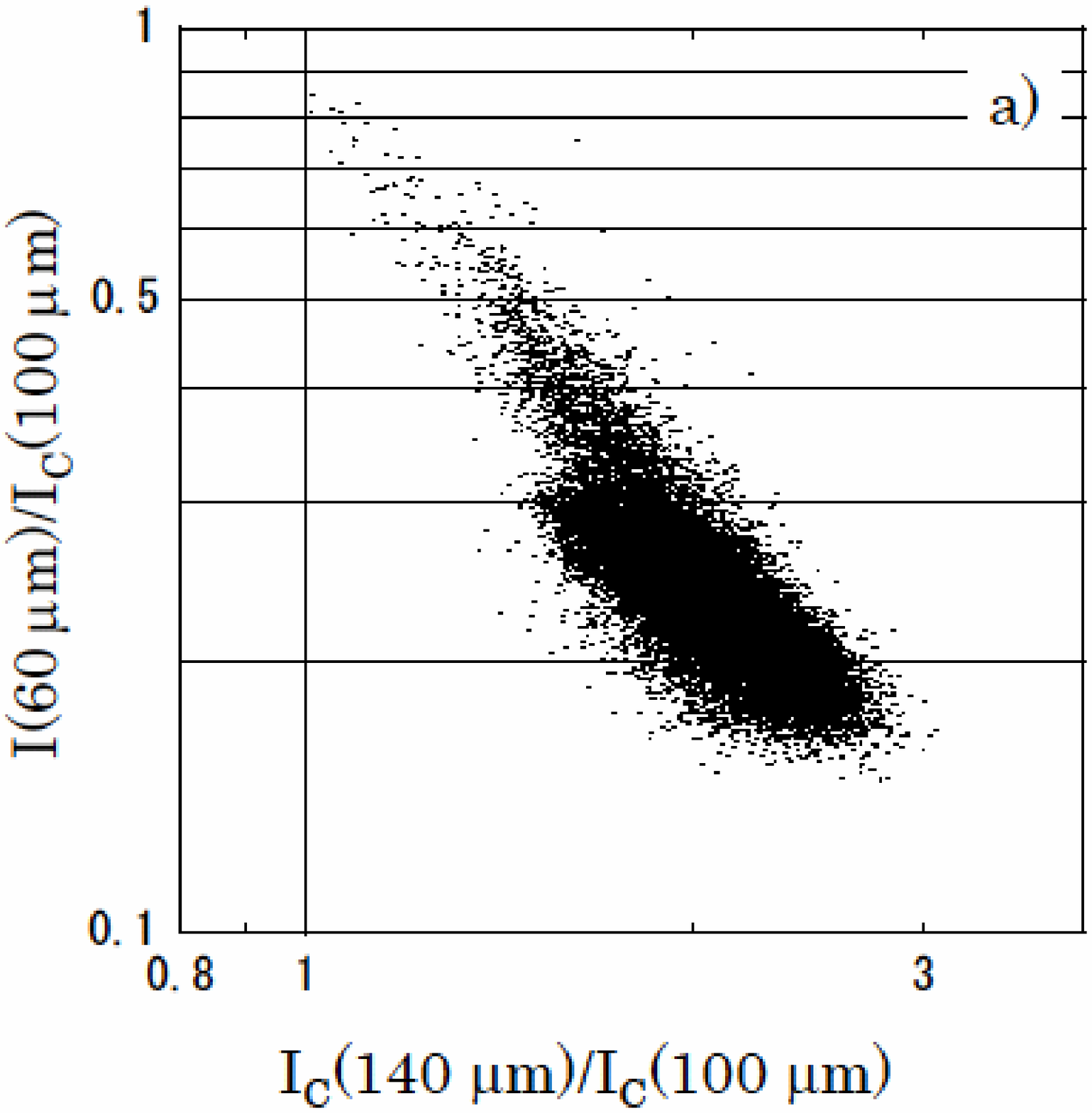}
  \end{center}
  \begin{center}
    \FigureFile(80mm,80mm){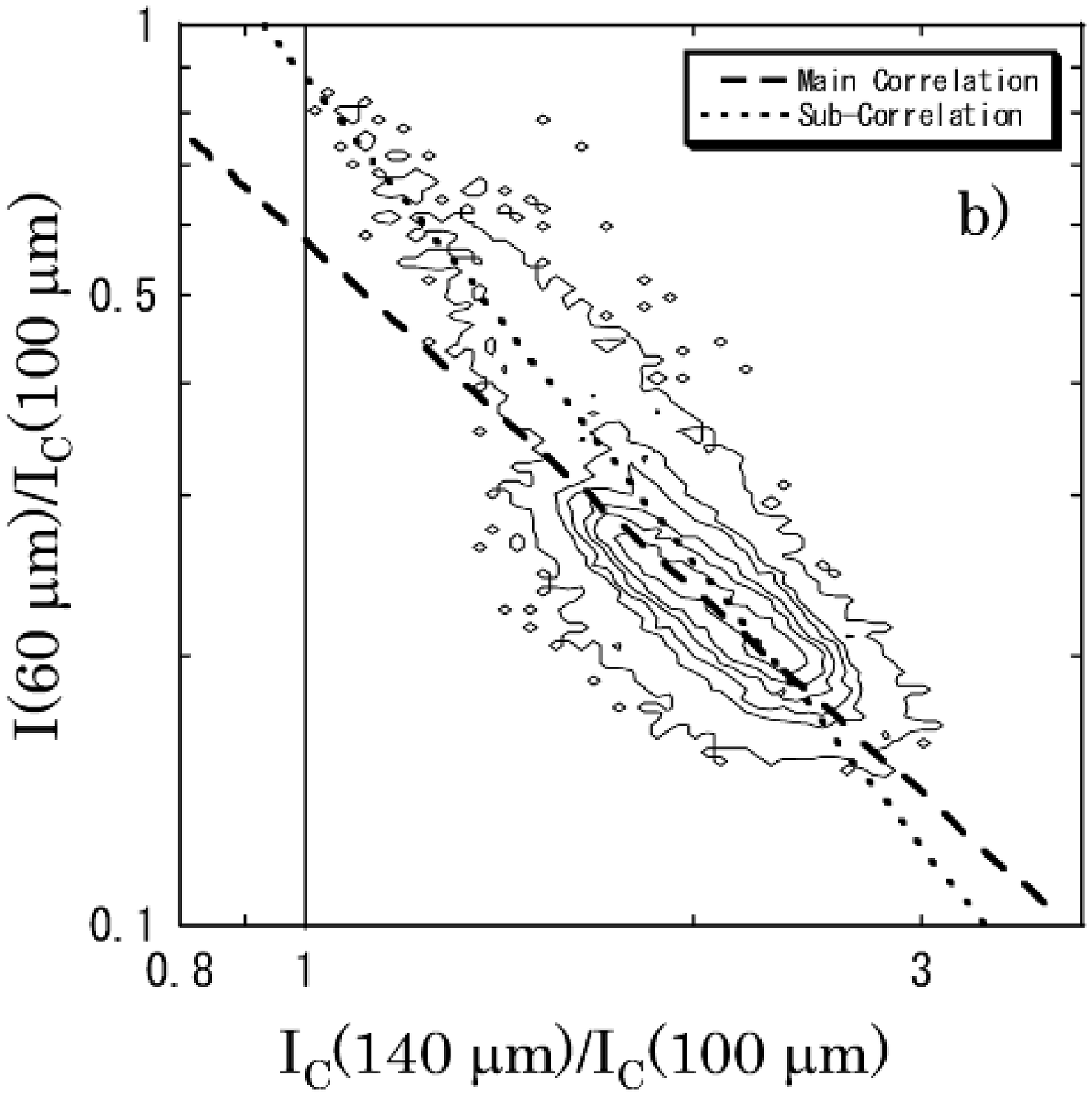}
  \end{center}
  \caption{Far-infrared color-color diagram of the Galactic plane. The relation between 
           [\(I_{C}(140\ \mu {\rm m})/I_{C}(100\ \mu {\rm m})\)] and [\(I(60\ \mu {\rm m})/I_{C}(100\ \mu {\rm m})\)] is presented. 
           {\bf a:} Plot of all the Galactic plane data. {\bf b:} Galactic plane data transformed to a contour map demonstrating the 
           density of data points on this diagram. The contour levels are set as 0.5, 0.5$N$, $N$, 2$N$, 3$N$ and 5$N$ data points 
           ($N$ is the average number of data points per box defined in the text). The dashed and the dotted lines indicate the main and 
           sub-correlations defined in the text, respectively.}
\end{figure}

In addition to the contour in figure 1b, all the data in the LMC and the SMC regions are plotted in figure 2. As mentioned in 
section 2.1, for the Galactic plane, we analyze only the region where the 10\ GHz radio continuum map is available in Handa et al. (1987). 
For the subsequent discussion, we divide the region at threshold antenna temperature of 0.5\ K. We present two color-color diagrams in 
figures 3a and 3b for the Galactic regions where the antenna temperature is higher and lower than 0.5\ K, respectively. In figure 3a, we 
also plot the data of the LMC 30Dor region, since the 30Dor region is one of the most active star-forming region in the Local Group and 
there seems to be a great variation of ISRF within a small area (e.g., Roberts \& Yusef-Zadeh 2005). In figure 3b, we also show the data of 
the LMC regions except for the 30Dor and the SMC region. These regions seem to have more moderate variation of ISRF than the 30Dor region.

Figures 1, 2 and 3 show that the Galactic plane data are concentrated around the dashed line shown in figure 1b. The tight relation defined 
by this line is called ``main correlation''. Besides, we see the other correlation sequence composed of a smaller number of data as shown 
by the dotted line in figure 1b. We call this correlation ``sub-correlation''. Comparing figures 3a and 3b, it is clear that the regions 
presenting the sub-correlation in color-color diagram correspond to the regions with high radio continuum intensity. The LMC data points 
except the 30Dor region data are concentrated on the upper left extension of the main correlation, while the correlation of the 30Dor data 
seem to be distributed between the main and the sub-correlation. Finally, the SMC data share the same region on the color-color diagram as 
the LMC (excluding the 30Dor) data.

\begin{figure}
  \begin{center}
    \FigureFile(80mm,80mm){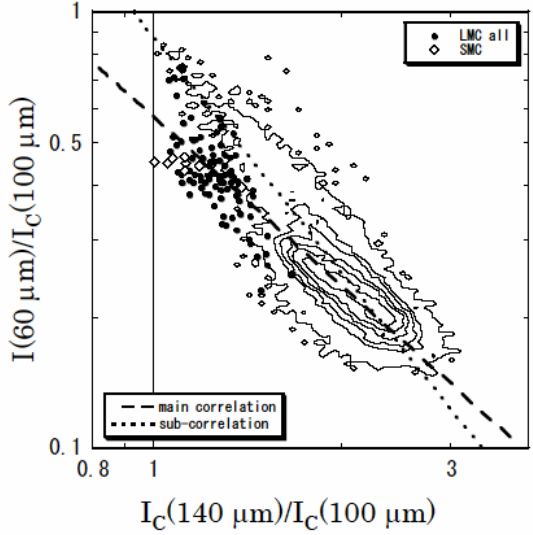}
  \end{center}
  \caption{Far-infrared color-color diagram of the Galactic plane, LMC and SMC from the COBE/DIRBE ZSMA map data. The LMC and the SMC data 
           are ploted by the filled circles and open diamondos, respectively. The contour show the distribution of the Galactic plane data 
           (same as figure 1b). The dashed line and the dotted lines indicate the main correlation and the sub-correlation, respectively.}
\end{figure}

\begin{figure}
  \begin{center}
    \FigureFile(80mm,80mm){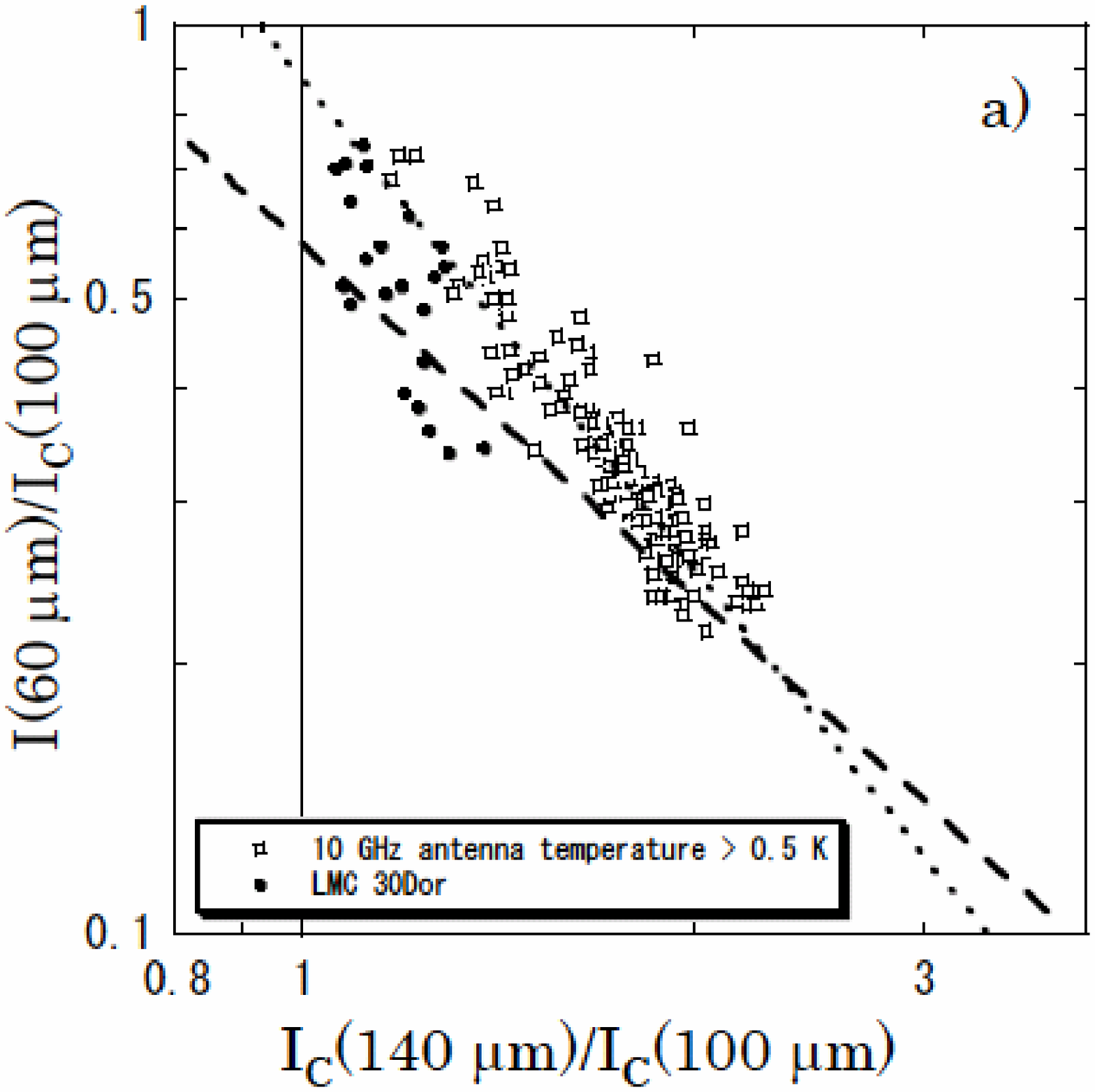}
  \end{center}
  \begin{center}
    \FigureFile(80mm,80mm){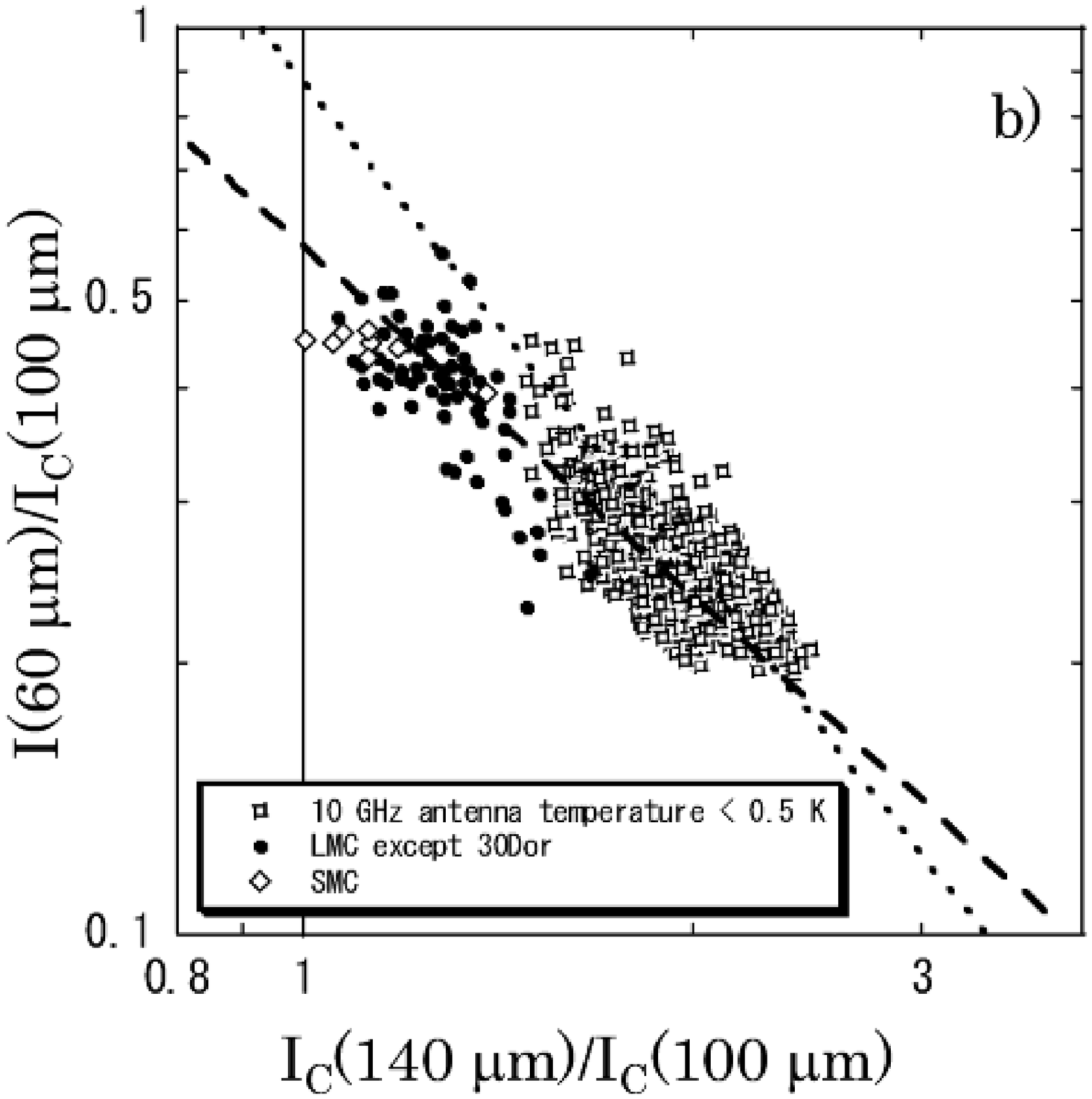}
  \end{center}
  \caption{{\bf a:} Color-color plot of the Galactic plane where the 10\ GHz radio continuum antenna temperature is higher than 0.5\ K (open 
           squares), and of the 30Dor data (filled circles). {\bf b:} Color-color plot of the Galactic plane where the 10\ GHz radio 
           continuum antenna temperature is lower than 0.5\ K (open squares). The filled circles and the open diamonds represent the LMC 
           data (30Dor excluded) and the SMC data, respectively.The dashed line and the dotted line indicate the main correlation and the 
           sub-correlation, respectively. }
\end{figure}

\subsection{Analysis Validity}

The comparison of our analysis with several other studies is presented in figure 4.

For the far-infrared radiation from the Galactic interstellar dust, the work by Arendt et al. (1998) is available. They identified the 
interstellar far-infrared radiation from $\vert b \vert > \timeform{25D.0}$ in the COBE/DIRBE data. Their result is shown by the filled 
square in figure 4. From figure 4, the far-infrared color of high Galactic latitude shown by Arendt et al. (1998) seems to be different 
from that of the Galactic plane. Since the analyzed region of Arendt et al. is different from ours, the discrepancy could be due to the 
difference in regions.

Sakon et al. (2004) investigated the far-infrared radiation from the Galactic plane by using the COBE data in the same regions of observed 
by the IRTS (InfraRed Telescope in Space; Murakami et al. 1996). They analyzed the data of four narrow areas of \(\vert b \vert < 
\timeform{4D}\) with various Galactic longitudes; area I (\(\timeform{-12D} \leq l \leq \timeform{-4D}\)), area\ II (\(\timeform{44D} \leq l 
\leq \timeform{52D}\)), area\ III (\(\timeform{-136D} \leq l \leq \timeform{-128D}\)), and area\ IV (\(\timeform{168D} \leq l \leq 
\timeform{176D}\)). They mentioned that there are two color correlations in far-infrared in two areas; area I and area II. They concluded 
that one color correlation is in the region of \(\vert b \vert < \timeform{1D.2}\) and they called this correlation Sequence A. The other 
correlation is found in the region of \(\vert b \vert > \timeform{1D.2}\), and is called Sequence B. Since the analyzed region of Sakon et 
al. (2004) overlaps with our region, we can compare their result with ours directly. 

It seems that the Sequences A and B in Sakon et al. (2004) agree to the sub- and the main correlations of our analysis, respectively. We have 
shown that the sub- and the main correlation are in high and low radio continuum regions, respectively (section 3.1). But they concluded that 
the Sequence A is found in \(\vert b \vert < \timeform{1D.2}\) while the Sequence B is seen in \(\vert b \vert > \timeform{1D.2}\). This 
difference could be caused by the following reason. From the radio continuum map presented by Handa et al. (1987), high radio continuum 
regions are concentrated on the region of \(\vert b \vert < \timeform{1D.0}\). Consequently, the analysis by Sakon et al. indicated that 
Sequence A is in the region of \(\vert b \vert < \timeform{1D.2}\). From our analysis, also the main correlation is found in the region of 
\(\vert b \vert < \timeform{1D.0}\). The absence of the Sequence A in \(\vert b \vert > \timeform{1D.2}\) region in Sakon et al. (2004) may 
be understood by the small probability that a line of sight contains active star-forming regions, because such regions tend to be compact. 
Actually, Sakon et al. did not find the Sequence A in the rest of two areas; area III and area IV.

For the LMC, there is no work which treats bands shorter and longer than 100\ $\mu$m at the same time. Therefore we refer to Aguirre et al. 
(2003) which determined the LG temperature in the LMC with only the COBE/DIRBE bands data longer than 100\ $\mu$m. They made use of the 
sub-mm to mm wavelength data taken by the balloon-borne telescope TopHat (Silverberg et al. 2005) and they observed near the celestial South 
Pole in four frequency bands, 245\ GHz, 400\ GHz, 460\ GHz and 630\ GHz. They summed up the flux in each band, treating the entire LMC region 
except 30Dor and the 30Dor region separately. For explaining observational results, they assume a multi-temperature model which is 
constructed by the mixing ratio proportional to $T^{\alpha}$ from a minimum ($=$ 2.7\ K) to a maximum LG temperature of dust. They estimated 
the maximum LG temperature of the entire LMC region except 30Dor as 26 K, and that of the 30Dor region as 27.5 K. The diameter of the 30Dor 
region defined by them is same to that defined by us and the center of their 30Dor region is about 
($\alpha$, $\delta$) = (\timeform{5h39m30s}, \timeform{-69D}).

The maximum LG temperature of the entire LMC region excluding the 30Dor is estimated at 23.8 $\pm$0.7 K. That of the 30Dor is estimated at 
23.8 $\pm$0.7 K, too. These maximum LG temperatures are derived from the highest temperature pixel of each region. These errors of LG 
temperatures are derived from the errors in the flux densities in the 100\ $\mu$m and 140\ $\mu$m band of COBE/DIRBE which are 10 and 5\ \% 
uncertainties, respectively. Those temperatures are lower than those given by Aguirre et al. (2003). These discrepancies are caused by the 
differences of calculation method of the maximum LG temperature. We assume one temperature component model for each pixel while they 
consider a continuous temperature distribution in the entire region.

Stanimirovic et al. (2000) investigated the far-infrared radiation from the SMC. Unlike our analysis, they adopted the entire region in the 
SMC. Aguirre et al. (2003) also observed the SMC region in the sub-mm to mm wavelength and determined the maximum LG temperature as 29 K. 
From figure 4, as for the SMC far-infrared color of Stanimirovic et al. (2000), \(I(60\ \mu {\rm m})/I_{C}(100\ \mu {\rm m})\) seem to be 
larger and \(I_{C}(140\ \mu {\rm m})/I_{C}(100\ \mu {\rm m})\) seem to be smaller than our results. But considering the error bar, our 
results are not incompatible with theirs. We estimate the maximum LG temperature in all the SMC regions as $24.6^{+0.7}_{-0.8}$ K. We 
propose the same reasons for the difference between our analysis and Aguirre et al.'s as shown in the previous paragraph.

\begin{figure}
  \begin{center}
    \FigureFile(80mm,80mm){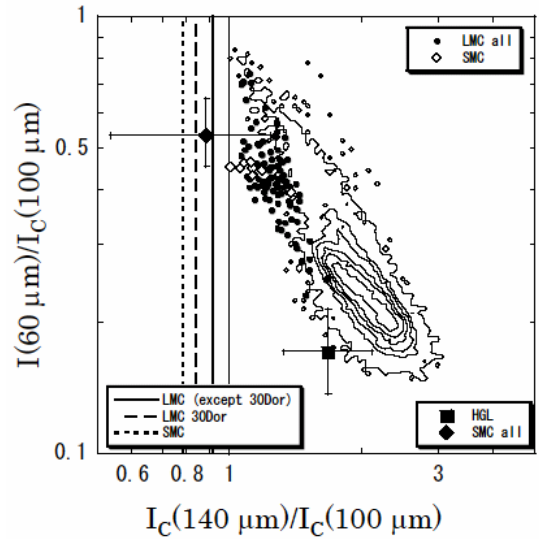}
  \end{center}
  \caption{Same as figure 2 for the Galactic plane (contours), LMC (filled circles) and SMC (open diamonds) data. The filled square is the 
           far-infrared color of High Galactic Latitude ($\vert b \vert > \timeform{25D.0}$) from Arendt et al. (1998) and a filled diamond 
           is the color of SMC from Stanimirovic et al. (2000), respectively. Solid line, dotted line and dashed line express the LMC 
           maximum LG temperature except 30Dor region, the LMC 30Dor maximum LG temperature and the SMC maximum LG temperature, respectively 
           from Aguirre et al. (2003).}
\end{figure}

\subsection{Comparison with Other Studies on Far-infrared Color Correlation}

We compare our results with two representative works on far-infrared SED of the interstellar dust (Li \& Draine 2002; Nagata et al. 2002). 
In figure 5, the color relations derived by those works are overplotted on our data in figure 4. We also show the color correlation of the 
``LG model'' which assumes that the interstellar dust consists of single-temperature LGs with an emissivity index of 2.

One of the studies on far-infrared SED is a series of papers written by Draine and Li (Draine \& Li 2001; Li \& Draine 2001; Li \& 
Draine 2002). They modeled the optical properties and the size distributions of interstellar grains and aimed to reproduce the extinction 
curve and near- to far-infrared SED of interstellar dust illuminated by an ISRF. To verify the validity of their dust model, they 
used the data of Arendt et al. (1998) in $\vert b \vert > \timeform{25D}$ as a template. Li \& Draine (2002) compared their model with 
observational near- to far-infrared SED of the SMC in Stanimirovic et al. (2000).

The other work on this issue is Nagata et al. (2002). One of their purposes is to estimate the total infrared intensity of individual 
galaxies by using the far-infrared data at $\lambda < 100\ \mu$m. They based their analysis on the far-infrared radiation 
from the Galactic plane. As a result, they found an empirical formula to estimate the total infrared intensity only from 60\ $\mu$m and 
100\ $\mu$m fluxes. They mentioned that this formula is more accurate than that of Helou et al. (1988) especially when the large grain 
temperature $T_{{\rm LG}}$ is lower than 25\ K.

As we see in figure 5, the far-infrared color in the Galactic plane cannot be explained by the LG model. This supports the contribution 
from VSGs to the 60\ $\mu$m band. Thus, if the LG temperature is determined from the 60\ $\mu$m and 100\ $\mu$m intensities (e.g., 
Kuiper et al. 1987), it is overestimated. We also observe from figure 5 that the color correlation by Li \& Draine (2002) is in good 
agreement with that in the high Galactic latitude by Arendt et al. (1998) and that in the SMC by Stanimirovic et al. (2000). But it agrees 
neither with the main correlation of the Galactic plane nor with the correlation of the LMC. The color correlation by Li \& Draine deviates 
downward from the main correlation in the color-color diagram. The color correlation by Nagata et al. agrees with the sub-correlation 
identified by us.

\begin{figure}
  \begin{center}
    \FigureFile(80mm,80mm){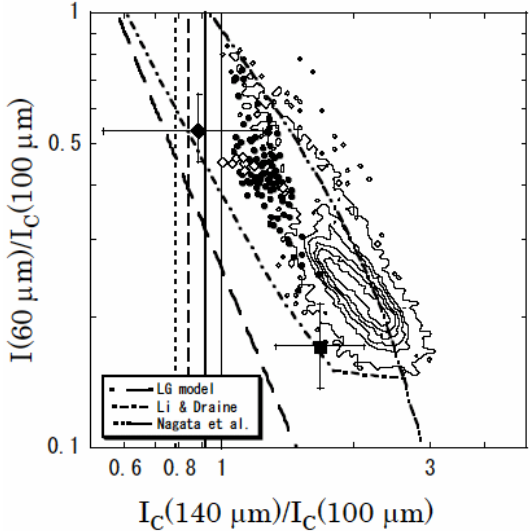}
  \end{center}
  \caption{Far-infrared color-color correlation predicted by the LG model, Li \& Draine's model and Nagata et al.'s model (long-dashed, 
           dot-dashed, dot-dot-dashed lines, respectively). The same observational data in figure 4 are also shown.}
\end{figure}

\section{Discussion}
\subsection{Far-Infrared Color Correlation}

We have found that more than 90 percent of the Galactic plane region shows a single color correlation in the far-infrared color-color 
diagram (figure 1). This tight correlation implies that the optical properties of interstellar grains in far-infrared do not vary or 
vary smoothly as a function of the ISRF intensity over a wide area in the Galactic plane. Moreover the ``overlap effects'' have little 
influence on the far-infrared color, that is, the far-infrared colors are dominated by interstellar dust illuminated by a uniform ISRF in a 
line of sight.

On the other hand, 5 percent of the Galactic plane has another color correlation. As shown in section 3.1, directions showing the 
sub-correlation agree with the directions with strong radio continuum. This implies that the sub-correlation should be caused by the overlap 
effect as explained in section 2.1. In section 4.3, we demonstrate how much the overlap effect changes far-infrared color correlation.

The LMC and SMC show significantly lower \(I_{C}(140\ \mu {\rm m})/I_{C}(100\ \mu {\rm m})\) than the Galactic plane. This indicates 
that the LG temperature is higher in the LMC and SMC than in the Galaxy. We propose some reasons for this. In the LMC and SMC, because 
star formation activities are higher than in the Galaxy, the mean ISRF intensity and dust temperature are higher. Low metallicity of 
the LMC and SMC may also be responsible for their large ISRF, since dust extinction tends to be small in metal-poor galaxies (Hirashita 
et al. 2001; Boissier et al. 2004).

In the LMC and SMC, smaller dust extinction may also cause a harder ISRF than in the Galaxy. When ISRF become harder such as in the LMC and 
SMC, even if the total intensity of ISRF is same, the ratio of UV photons to all photons become larger and the probability that UV 
photons hit VSGs become larger. As a result, typical VSG temperature becomes higher and the 60\ $\mu$m intensity becomes higher than in the 
Galaxy. However, the far-infrared color correlations of the LMC and SMC seem to be on the main correlation of the Galactic plane. If the 
main correlation exhibits the general optical properties of interstellar dust, this may indicate that there is little difference in the 
hardness of ISRF among the Galaxy, the LMC and the SMC. This ISRF hardness problem will be investigated in our future work. 

On the other hand, the color correlation in the 30Dor region is similar to the sub-correlation of the Galaxy. This suggests that the overlap 
effect is important in the 30Dor (see section 4.3).

From Shibai, Okumura \& Onaka (1999), the mean value of \(I_{C}(140\ \mu {\rm m})/I_{C}(100\ \mu {\rm m})\) in the range of $\vert l \vert 
<$ \timeform{90D} ($\vert b \vert <$ \timeform{5D}) is lower than the mean value in the range of $\vert l \vert >$ \timeform{90D} ($\vert b 
\vert <$ \timeform{5D}) . They concluded that the variation of the ISRF is not very large. To confirm this fact and investigate the 
variation of the far-infrared color in the Galactic plane depending on the Galactic longitude, we divide the COBE/DIRBE Galactic plane data 
into every $\vert l \vert$.

In figure 6, we show the color-color diagrams in various ranges in the Galactic longitudes. The longitudes are divided in a width of
\timeform{30D}. Used data is same as figure 1a and 1b. From these figures, components around the dashed line shift along the main 
correlation as the Galactic longitudes change. This means that the dust temperature continuously drops as the direction changes from the 
Galactic center to the anti-center. Since the dust temperature reflect the ISRF, the trend is consistent with the picture that the ISRF is a 
decreasing function of the Galactocentric distance. 

We calculate the mean value of \(I_{C}(140\ \mu {\rm m})/I_{C}(100\ \mu {\rm m})\) of the each Galactic longitude range from the value of 
all pixels of the each range. The mean LG temperatures in each range are derived from the mean value of 
\(I_{C}(140\ \mu {\rm m})/I_{C}(100\ \mu {\rm m})\) in the each range. The mean LG temperature in $\vert l \vert <$ \timeform{30D} is 18.2 K, 
while that in $\vert l \vert >$ \timeform{150D} is 16.3 K; therefore, the mean ISRF intensity in the direction of the Galactic center is 
twice as high as in the direction of the Galactic anti-center by using equation (A.3).

The present interpretation about the far-infrared color correlations is that the main correlation appears in the directions of no strong UV 
sources in a line of sight, and the sub-correlation appears in the directions of some strong UV sources. From figure 6 and the comparison 
with the radio intensity map (above-mentioned), this interpretation is most plausible.

\begin{figure*}
  \begin{center}
    \FigureFile(120mm,180mm){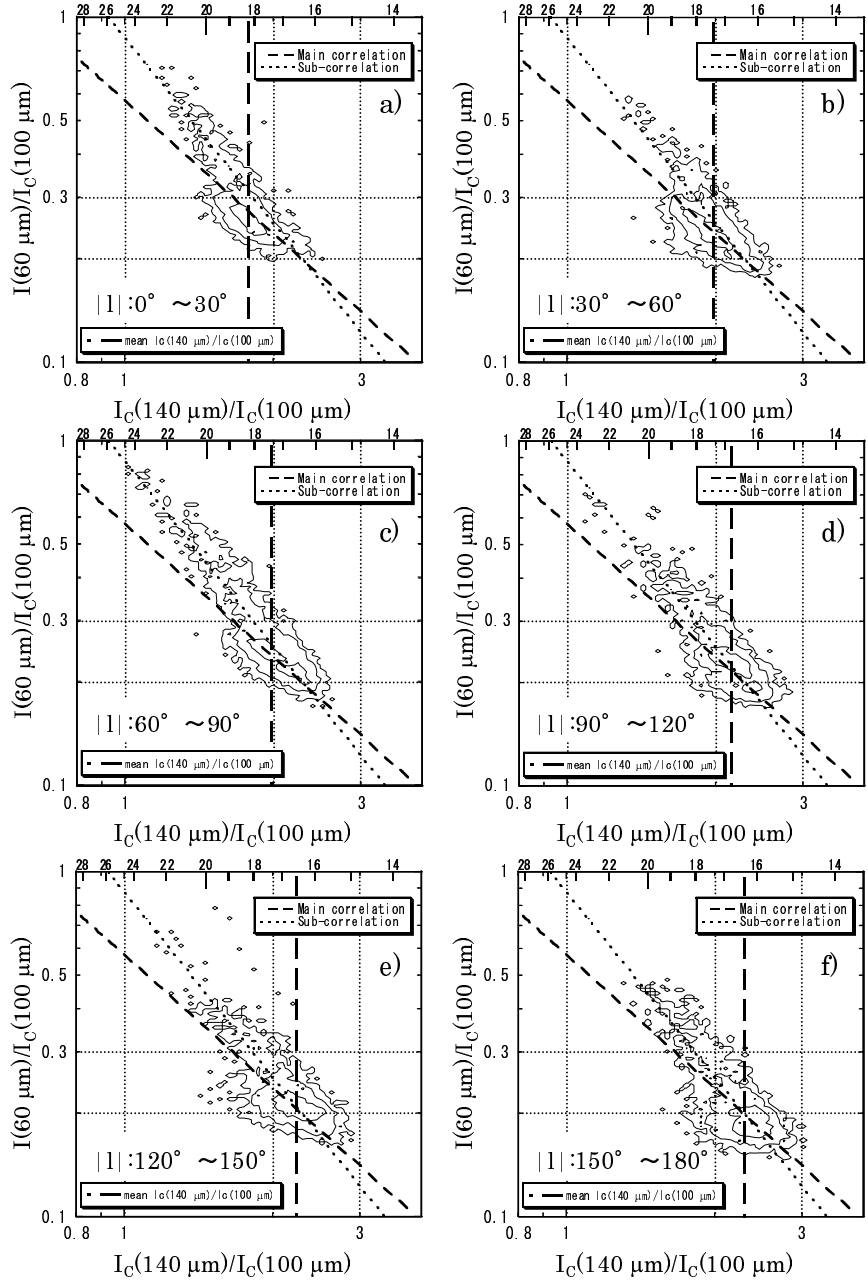}
  \end{center}
  \caption{Far-infrared color-color diagram of the Galactic plane in each range of $\vert l \vert$. The same data as figure 1 are used. They 
           are transformed to contour maps. The contour levels are set as 0.5, 0.5$N$, and 4$N$. We show the main and sub-correlation on 
           each panel. We also show the mean value of \(I_{C}(140\ \mu {\rm m})/I_{C}(100\ \mu {\rm m})\) of each region by thick vertical 
           long-dashed lines. Numbers on the upper horizontal axis are LG temperatures derived from the value of \(I_{C}(140\ \mu 
           {\rm m})/I_{C}(100\ \mu {\rm m})\) and the SED shape of \(\nu^{2}B_{\nu}(T)\). The Galactic longitude range and the $N$ value of 
           each panel are {\bf a:} \timeform{0D.0} $< \vert l \vert <$ \timeform{30D.0} and $N =$ 14, {\bf b:} \timeform{30D.0} $< \vert l 
           \vert <$ \timeform{60D.0} and $N =$ 12, {\bf c:} \timeform{60D.0} $< \vert l \vert <$ \timeform{90D.0} and $N =$ 10, {\bf d:} 
           \timeform{90D.0} $< \vert l \vert <$ \timeform{120D.0} and $N =$ 10, {\bf e:} \timeform{120D.0} $< \vert l \vert <$ 
           \timeform{150D.0} and $N =$ 9, and {\bf f:} \timeform{150D.0} $< \vert l \vert <$ \timeform{180D.0} and $N =$ 7.5.}
\end{figure*}

\subsection{Origin of the Main Correlation}

As shown in figure 5, the main correlation is not explained by the LG model, which indicates that other dust components such as VSGs 
should be considered. However, the main correlation does not agree with Li \& Draine (2001), although they include VSG components. On the 
other hand, the correlation identified by Nagata et al. (2002) is the sub-correlation defined by us. Thus we do not have any appropriate 
physicals model which explains the main correlation. Shibai et al. (1999) took into account the two-photon heating to explain the 
far-infrared color correlation of the Galactic plane, and fitted their model to the same sub-correlation as we identified. However, the 
two-photon heating model seems to be marginally consistent with the main correlation as well as with the sub-correlation, especially in 
the low $T_{{\rm LG}}$ (i.e., low $G_{{\rm UV}}$ in Appendix) regime. See Appendix for the color correlation predicted by the two-photon 
heating model.

\subsection{Origin of the Sub-Correlation}

The sub-correlation may result from the ``overlap effect'', i.e., effect of observing various interstellar grains illuminated by various 
ISRF in a line of sight (section 3.2). As shown in figures 3a and 3b, the sub-correlation is seen in the directions with strong radio 
continuum, where the overlap effect is considered to be large (section 2.1). The other reason explaining origin of the sub-correlation, 
the difference of ISRF hardness is available. From our estimation by using Draine \& Li interstellar dust model, the main and 
sub-correlations cannot be explained by changing only the intensity of ISRF. To reproduce the sub-correlation by using Draine \& Li dust 
model, we shall have to change the size distribution of interstellar dust and the hardness of ISRF at the same time. 

To examine the overlap effect, we construct the following model. For simplicity, we consider mixture of only two components. We assume 
that the first component has a common color, 
(\(I_{C}(140\ \mu {\rm m})/I_{C}(100\ \mu {\rm m})\), \(I(60\ \mu {\rm m})/I_{C}(100\ \mu {\rm m})\)) = (2.50, 0.176), which is typical of 
the Galactic plane in the region of \(\vert l \vert > \timeform{90D.0}\) (Shibai et al. 1999). We examine three colors for the second 
component: (\(I_{C}(140\ \mu {\rm m})/I_{C}(100\ \mu {\rm m})\), \(I(60\ \mu {\rm m})/I_{C}(100\ \mu {\rm m})\)) = (1.50, 0.337), (1.00, 
0.570) and (0.500, 1.42), called Cases 1, 2, and 3, respectively. The first two colors of the second components follow the main correlation, 
and the last color (\(I_{C}(140\ \mu {\rm m})/I_{C}(100\ \mu {\rm m})\), \(I(60\ \mu {\rm m})/I_{C}(100\ \mu {\rm m})\)) = (0.50, 1.42) 
is introduced based on the LG model to represent hot dust in high ISRF environments. In each Case, we change the mixing ratio of the first 
component to the second component from 0 to 100. The color relations calculated by these mixtures are plotted in figure 7 and compared 
with the observational data of the Galactic plane (contour in figure 7).

From figure 7, we observe that the calculated color correlations are located near to the sub-correlation in the color-color diagram. The 
fact that the color correlation shifts above the main correlation by the mixing effects can qualitatively explain the position of the 
sub-correlation. But it is difficult to reproduce the overall sub-correlation by a simple mixture of two components, especially at low 
\(I_{C}(140\ \mu {\rm m})/I_{C}(100\ \mu {\rm m})\). It may be necessary to consider an elaborate model of radiation transfer effects. 
For example, Onaka et al (2003) demonstrated how far-infrared SED are changed by varying the distribution function of ISRF intensity.

\begin{figure}
  \begin{center}
    \FigureFile(80mm,80mm){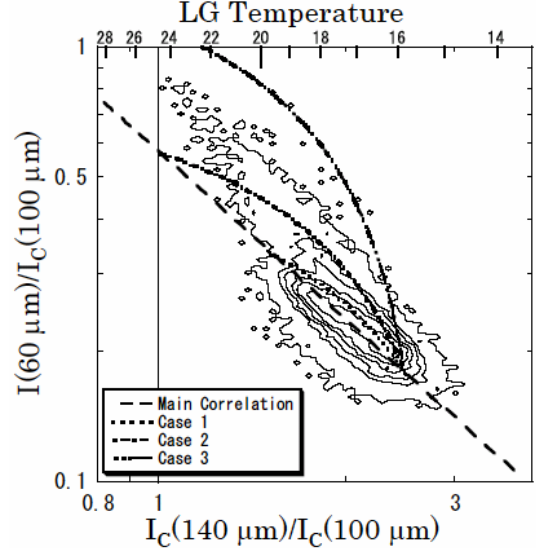}
  \end{center}
  \caption{Color correlation in Case 1, 2 and 3 for the two component mixture (dotted, dot-dashed and dot-dot-dashed lines, respectively). 
           The dashed line represents the main correlation, and the contour shows the distribution of the Galactic plane data (figure 1b).
           Numbers on the upper horizontal axis are LG temperatures derived from the value of (\(I_{C}(140\ \mu {\rm m})/I_{C}(100\ \mu 
           {\rm m})\) and the shape of \(\nu^{2}B_{\nu}(T)\).}
\end{figure}

\subsection{Comparison with Nearby Galaxies}

The main correlation can be fitted as
\begin{equation}
 \frac{I_{\nu}(140\ \mu {\rm m})}{I_{\nu}(100\ \mu {\rm m})}
                                                   =0.65 \left( \frac{I_{\nu}(60\ \mu {\rm m})}{I_{\nu}(100\ \mu {\rm m})} \right)^{-0.78}
\end{equation}
We compare this relation with the colors of nearby galaxies compiled by Nagata et al. (2002). For this aim, we derive $T_{{\rm LG}}$ from 
\(I_{C}(140\ \mu {\rm m})/I_{C}(100\ \mu {\rm m})\) by fitting the Planck function multiplied by $\nu^{2}$. Then we finally obtain the 
relation between \(I(60\ \mu {\rm m})/I_{C}(100\ \mu {\rm m})\) and $T_{LG}$ for the main correlation as shown in figure 8. In the same 
figure, we also present the relations calculated from the LG model, and from the results in Li \& Draine (2001) and in Nagata et al. (2002). 
From figure 8, we conclude that the main correlation reproduces the far-infrared color correlation of nearby galaxies.

Dale \& Helou (2002) modeled typical infrared SEDs of nearby galaxies as a single parameter family of 60 -- 100\ $\mu$m flux ratio. In 
figure 8, we show the color correlation found in their work. Dale \& Helou color correlation is quite similar to the main correlation of 
the present work. 

\begin{figure}
  \begin{center}
    \FigureFile(80mm,80mm){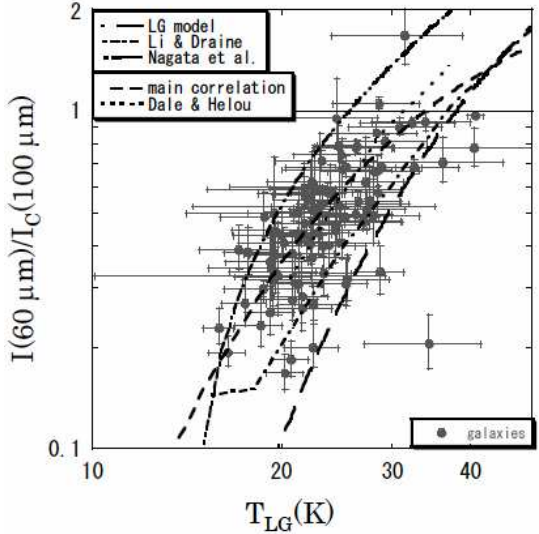}
  \end{center}
  \caption{Relation between the LG temperature and the 60 -- 100\ $\mu$m flux ratio. The nearby galaxies listed in Nagata et al. (2002) 
           are shown by gray filled circles with error bars. The long-dashed, dot-dashed and dot-dot-dashed lines show the color relations 
           calculated by the LG model, Li \& Draine's model, Nagata et al.'s model, respectively. The main correlation and the color 
           correlation of Dale \& Helou (2002) are expressed by the short-dashed and dotted lines, respectively.}
\end{figure}

\section{Conclusion}

We have investigated the far-infrared SED of the Galactic plane, the LMC and the SMC with the COBE/DIRBE ZSMA map at $\lambda =$ 60\ $\mu$m, 
100\ $\mu$m and 140\ $\mu$m. For the Galactic plane, we have adopted the region with \(\vert b \vert < \timeform{5D.0}\). To avoid an overlap 
effect (effect of detecting all the cumulative far-infrared radiation from interstellar dust illuminated by various of ISRF intensities in 
a line of sight), we have compared the Galactic far-infrared map with the 10\ GHz radio continuum map and have 
identified high radio continuum regions as regions with strong overlap effect. We have also analyzed the LMC and SMC regions, where the 
contribution from the Galaxy has been subtracted. To avoid the overlap effect, the 30Dor region has been separated from the other LMC area.

We have examined the far-infrared color-color (60\ $\mu$m --- 100\ $\mu$m vs. 140\ $\mu$m --- 100\ $\mu$m) diagram of the Galactic plane, 
the LMC and the SMC. We have found a common far-infrared color correlation (called ``main correlation'' in this paper) among those three 
objects. This fact indicates that there is a common far-infrared optical property in these three galaxies in spite of difference in ISRF 
intensity and metallicity. Then the main correlation reveals the existence of grain components other than LGs. Although other components 
such as VSGs are taken into account, currently available models by Li \& Draine (2001) fail to explain the main correlation. 

On the other hand, another color correlation (called ``sub-correlation'') is found in regions with high radio continuum. The sub-correlation 
itself is identified by Shibai, Okumura, \& Onaka (1999). The sub-correlation can be qualitatively explained by mixing two or more 
components, each of which have a color satisfying the main correlation or a color of a hot LG component (representative of active 
star-forming regions). Finally, we heve compared the main correlation with the color of nearby galaxies. We have found that the main 
correlation reproduces the far-infrared color correlation of nearby galaxies.

\medskip

{\bf Acknowledgement:} We thank A. Kawamura, J.P. Bernard, S. Kim, I. Sakon and T. Onaka for helpful discussions. R. P. Verma assisted in 
correction of this paper. H. Hirashita has been supported by University of Tsukuba Research Initiative.

\appendix
\section*{60\ $\mu$m Intensity Normalized to the Total Intensity of LGs}

Shibai, Okumura \& Onaka (1999) referred to Okumura (1998) and investigated the process of grain heating by comparing 60\ $\mu$m intensity 
normalized to the total LG intensity with Ultra-Violet (UV) ISRF intensity. They supposed that the radiation at $\lambda =$100\ $\mu$m and 
140\ $\mu$m originates from LGs which are in radiative equilibrium with ISRF. On the other hand, the radiation at $\lambda =$60\ $\mu$m is 
attributed to stochastically heated VSGs plus the LGs. The contribution of LGs to the 60\ $\mu$m intensity depends on the LG temperature, 
which is determined by fitting a spectral of \(\nu^{\beta}B_{\nu}(T)\) of to the intensities at $\lambda =$100\ $\mu$m and 140\ $\mu$m. 
Because the maximum of LG temperature is around 30 K in the analyzed region of their study, the contribution from LGs to the 60\ $\mu$m 
band is negligible, and almost all the radiation at $\lambda$=60\ $\mu$m comes from VSGs. Then the 60\ $\mu$m intensity normalized to the 
total intensity of LGs can be interpreted as the far-infrared intensity radiation from VSGs normalized to that from LGs. We estimate this 
normalized intensity as 
\begin{equation}
\frac{\nu I_{\nu}(60\ \mu {\rm m})}{I_{{\rm LG}}} = \frac{(5\ {\rm THz}) \times I(60\ \mu {\rm m})}
{\tau_{100\ \mu {\rm m}} \times \int_{0}^{\infty} \left(\frac{\nu}{3\ {\rm THz}}\right)^{2} B_{\nu}(T_{{\rm LG}})d\nu}
\end{equation}
where $\tau_{100\ \mu{\rm m}}$ is the optical depth at $\lambda =$100\ $\mu$m and $B_{\nu}(T)$ is the Planck function. The LG temperature 
$T_{{\rm LG}}$ is calculated from the 140\ $\mu$m -- 100\ $\mu$m intensity ratio. We convert $T_{{\rm LG}}$ into the UV ISRF in units of 
1.6 $\times$ 10$^{-3}$ erg cm$^{-2}$ s$^{-1}$ (Habing 1968), $G_{{\rm UV}}$, by using the interstellar dust emissivity model of Draine 
\& Lee (1984),
\begin{equation}
\frac{Q_{{\rm abs}}({{\rm UV}})}{Q_{{\rm abs}}(100\ \mu {\rm m})}=700
\end{equation}
obtaining
\begin{equation}
G_{{\rm UV}}=1.84 \times 10^{-7}T_{{\rm LG}}{}^{6}
\end{equation}
where $Q_{{\rm abs}}({{\rm UV}})$ and $Q_{{\rm abs}}(100\ \mu{\rm m})$ are the absorption efficiencies of grains at UV and 
$\lambda =$100\ $\mu$m, respectively.

The dependence of $\nu I_{\nu}(60\ \mu{\rm m})/I_{{\rm LG}}$ on the ISRF intensity can be used to constrain the heating mechanism of VSGs. 
If $\nu I_{\nu}(60\ \mu{\rm m})/I_{{\rm LG}} \propto G_{{\rm UV}}^{n}$, $n$ photons are absorbed on a cooling timescale of a VSG (Okumura 
1998). Shibai, Okumura \& Onaka (1999) concluded that the heating process of VSGs in the Galaxy is two-photon heating. However, they 
identified only the sub-correlation (see figure 1b). We consider that the main correlation is more essential and that the sub-correlation 
arises from the overlap effect (see section 4.3).

We show the relation between $G_{{\rm UV}}$ and $\nu I_{\nu}(60\ \mu{\rm m})/I_{{\rm LG}}$ of the Galactic plane, the LMC and the SMC in 
figure 9. The contours, which show the density of data points on the diagram, are drawn in the same way as in section 3.1 but the logarithmic 
bin size is 0.2. We also show the relation calculated by the LG model, Li \& Draine (2002)'s model and Nagata et al. (2002)'s color 
correlation in the same figure. In addition, the slopes of $\nu I_{\nu}(60\ \mu{\rm m})/I_{{\rm LG}} \propto G_{{\rm UV}}^{0}$ and 
$\nu I_{\nu}(60\ \mu{\rm m})/I_{{\rm LG}} \propto G_{{\rm UV}}^{1}$ (one- and two-photon heating, respectively) are presented.

From figure 9, we see that the main correlation is not explained by the one-photon heating. The slope at $G_{{\rm UV}} \ltsim 10$ is rather 
similar to that predicted by the two-photon process. However, the main correlation at $G_{{\rm UV}} \gtsim 10$ cannot be explained by 
two-photon heating. In general, the two-photon heating becomes prominent if $G_{{\rm UV}}$ becomes large; thus, an increase, not a decreace 
of the slope is predicted as $G_{{\rm UV}}$ increases. Therefore, the transition from the one-photon heating to the two-photon heating fails 
to explain the change of the slope.

\begin{figure}
  \begin{center}
    \FigureFile(80mm,80mm){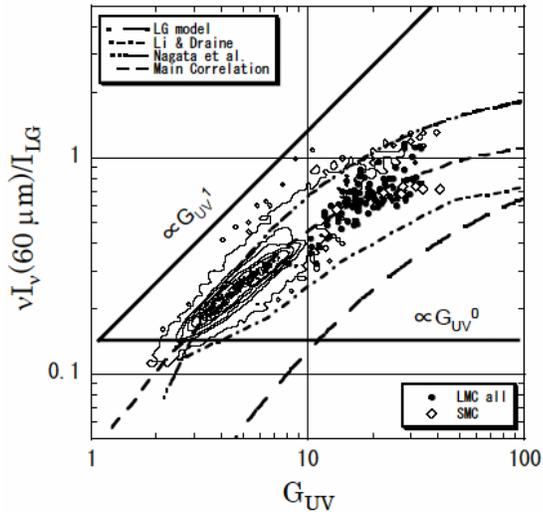}
  \end{center}
  \caption{Relation between the UV ISRF intensity and the 60\ $\mu$m intensity normalized to the total LGs intensity of the Galactic plane, 
           LMC and SMC. The Galactic plane data are expressed by the contour map demonstrating the degree of data point concentration, the 
           LMC and the SMC data are expressed by filled circles and open diamonds, respectively. The LG model, Li \& Draine's model, Nagata 
           et al.'s model and the main correlation are expressed by the long-dashed, dot-dashed, dot-dot-dashed and short-dashed lines, 
           respectively. Two solid lines which represent the slope of $\nu I_{\nu}(60\ \mu{\rm m})/I_{{\rm LG}} \propto G_{{\rm UV}}^{0}$ 
           and $\propto G_{{\rm UV}}^{1}$ are also shown.}
\end{figure}

\end{document}